\documentclass[reqno,12pt,a4]{amsart}
\usepackage{amsmath,amssymb,tabularx,setspace,color}
\usepackage{pdflscape}

\newtheorem{definition}{Definition}
\newtheorem{remark}{Remark}[section]

\usepackage{slashbox}
\usepackage{multirow}
\usepackage{hyperref}
\usepackage{multirow}
\usepackage[bf,small,tableposition=top]{caption}
\usepackage{amsthm}
\usepackage{lineno}
\usepackage[figuresright]{rotating}
\usepackage{enumerate}
\usepackage[T1]{fontenc}
\usepackage[utf8]{inputenc}
\usepackage{graphics}
\usepackage{graphicx}
\usepackage{lscape}
\usepackage[dvipsnames]{xcolor}
\usepackage{mathrsfs}
\usepackage{hyperref}
\usepackage{textcomp}
\usepackage{listings}
\usepackage{amsmath}
\usepackage[margin=3.5cm]{geometry}
\makeatletter
\renewcommand\section{\@startsection {section}{1}{\z@}%
	{-3.5ex \@plus -1ex \@minus -.2ex}%
	{2.3ex \@plus.2ex}%
	{\normalfont\large\bfseries}}
\makeatother

\newtheorem{Corollary}{Corollary}[section]

\begin{document}
\begin{singlespace}
	\begin{center}
		\begin{doublespace}
			{\Large\bf Proportional mean model for panel count data with multiple modes of recurrence }\\
		\end{doublespace}
		Sreedevi E. P.\footnote{ Corresponding Author Email: sreedeviep@gmail.com } and Sankaran P. G.$^{2}$ \\
	$^{1}$ SNGS College, Pattambi.\\
	$^{2}$ Cochin University of Science and Technology, Cochin.
\end{center}
\vspace{0.3cm}	
\textit{\textbf{Abstract}}: Panel count data is common when the study subjects are exposed to recurrent events, observed only at discrete time points. In this article, we consider the regression analysis of panel count data with multiple modes of recurrence. We propose a proportional mean model to estimate the effect of covariates on the underlying counting process
due to different modes of recurrence. The simultaneous estimation of baseline cumulative mean functions and regression parameters of $(k>1)$ recurrence modes are studied in detail. Asymptotic properties of the proposed estimators are also established.  A Monte Carlo simulation study is carried out to validate the finite sample behaviour of the proposed estimators. The methods are applied to a real data arising from skin cancer chemoprevention trial.\\


\end{singlespace}

\section{Introduction}	
In many longitudinal studies on recurrent events in lifetime data analysis, instead of observing the time to occurrence of event, we may only observe the number of events experienced by a subject in a given period of time. If each subject can be observed at more than one time points, the number of events between two successive observation times is available. The data obtained in this form is known as panel count data (Kalbfleisch and Lawless,1985; Sun, 2009). Panel count data frequently arise in many fields such as clinical trials, epidemiological studies and engineering, when continuous follow-up to obtain exact event times of each subject is infeasible or too costly (Chiou et al., 2019). Some authors refer panel count data as  interval count data or interval censored recurrent event data (Lawless and Zhan, 1998; Thall, 1988). An extensive review of panel count data is given in Sun and Zhao (2013). Note that when the subjects can be  observed only at a single time point, we obtain current status data which is explored in the monograph by Sun (2007).

The standard methods in the analysis of panel count data are focused on the mean function or the rate function of the underlying recurrent event process. An estimator for the mean function based on isotonic regression theory is developed by Sun and Kalbfleisch (1995). Wellner and Zhang (2000) discussed likelihood  based nonparametric estimation methods for the mean function and proposed a nonparametric maximum likelihood estimator (NPMLE) and a nonparametric maximum pseudo likelihood estimator (NPMPLE) for the same.  Wellner and Zhang (2000) also studied the asymptotic properties of both NPMPLE and NPMLE. Thall and Lachin (1988) and Lawless and Zhan (1998) considered the analysis of panel count data using rate functions. Some of the recent developments in the analysis of panel count data include  Xu et al. (2017) and Chiou et al. (2018) among others.

In panel count data, it is common to observe a covaraite vector $Z$ for each subject which affect the underlying counting process of recurrent events. Two different approaches employed for the analysis of regression models for panel count data are either by using maximum likelihood methods or by applying the generalized estimating equation approach.  Some important developments in this area include Sun and Wei (2000), Wellner and Zhang (2007), Zhang (2002) and Hu et al. (2003). Regression analysis of panel count data with informative observation times is considered by Haung et al. (2006), Sun et al. (2007) and Zhao and Tong (2011). Covaraites with measurement error for panel count data was studied by Kim (2007). Recently Chiou et al. (2019) reviewed various semiparametric regression modelling approaches for panel count data using R programming language.

When the study subjects are exposed to recurrent events of several types, we observe the recurrence due to each possible mode (cause) of recurrence at different observation times.  As a result, we obtain  panel count data with multiple failure modes. For example consider the data on skin cancer chemoprevention trial discussed in Sun and Zhao (2013). The cancer recurrences of 290 patients with a history of non-melanoma skin cancers are observed at different monitoring times. The types of cancers are classified into Basal cell carcinoma and Squamous cell carcinoma and the recurrences due to both types of cancers at each monitoring time are observed for each individual. Covariate information on age, gender, number of prior tumours and DFMO status is also observed for each individual. Accordingly, we have panel count data with multiple modes of recurrence with covariates. A detailed analysis of the data is given in Section 5.

Even though panel count data was a topic of research interest from last two decades, only a sparse amount of literature is available on  panel count data with multiple failure modes.  Sreedevi and Sankaran (2020) developed an estimator for cause specific mean function and Sankaran et al. (2021) studied cause specific rate functions of the underlying recurrent event processes when subjects are exposed to more than one recurrence mode. Both these works considered data without covariates.  Regression analysis of panel count data with multiple failure modes is not studied yet. Motivated by this, in this article we propose a proportional mean model to estimate the regression parameters and baseline cumulative mean functions of panel count data exposed to more than one mode of recurrence.

The rest of the article is organized as follows. In Section 2, we propose a new proportional mean model to estimate the baseline cumulative mean functions and regression parameters due to each mode of recurrence simultaneously . A simple iterative algorithm is derived for the estimation. Asymptotic properties of the proposed estimators are established in Section 3. In Section 4, the finite sample behaviour of the proposed estimators is validated through a Monte Carlo simulation study. The proposed procedures are illustrated using a real data on skin cancer chemoprevention trial in Section 5. Finally, Section 6 gives concluding remarks with a discussion on possible future works.
\noindent
\section { The proportional mean model }
Consider a  study on $n$ individuals exposed to the recurrent events due to $  \{1,2,...,k\}$  different causes. Assume that the event process is observed only at a sequence of random monitoring times. Consequently, the counts of the event recurrences due to each mode in between the observation times are only available; the exact recurrence times remain unknown.

Define a counting process $ N_j(t)=\{N_j(t); t\ge 0\}$ where $N_j(t)$ denote the number of recurrences of the event due to cause $j$ upto time $t$. Now, $E(N_j(t))=\Lambda_{j}(t)$ for $ j=1,2,..,k$ denote the expected number of cumulative events due to cause (mode) $j$ upto time $t$. The function $ \Lambda_{j}(t) $ is the mean function of the counting process $N_j(t)$ and can be termed as the cause specific mean functions (Sreedevi and Sankaran, 2020). Assume that, corresponding to each subject we observe a $d \times 1$ vector of covariates denoted by $Z$. Our interest is to study $E(N_j(t) \vert Z )=\Lambda_{j}(t \vert Z)$ for $ j=1,2,..,k$, the expected number of cumulative events due to cause $j$ upto time $t$ conditionally on covariate vector $Z$. To estimate the effect of covariate vector $Z$ on lifetime $T$,
we propose the proportional mean model given by
\begin{equation}\label{pmm}
\Lambda_{j}(t\vert Z) = \Lambda_{0j}(t)\text{exp}(\beta_j^{'}Z)~~~~~j=1,2,...,k
\end{equation}
where $\Lambda_{0j}(.)$ is the completely unspecified baseline mean function and $\beta_j$ is the $ d \times 1 $ vector of regression parameters corresponding to cause $j$. When $k=1$, the model in Eq. (\ref{pmm})  reduces to the proportional mean model for panel count data with a single mode of recurrence studied by Sun and Wei (2000), Zhang (2002) and  Wellner and Zhang (2007).

Now, we discuss the structure of panel count data with covariates which is exposed to multiple modes of recurrence.  Let $M$ be an integer valued random variable denoting the number of observation times which may be different for each individual and $\underline{T} =\{T_{m,p}, p=1,2,...,m ; m=1,2,...\} $ be the set of observation times. Now  $T_{m,p-1} \le T_{m,p}$ for $p=1,2,...,m $ and for all possible values of $m$. Assume that $N_j(t)$ and $(M,\underline{T} )$ are independent. Let $N_{M,p}^{j}$ denote the number of recurrences of the event due to cause $j$ upto monitoring time $M$ for $p=1,2,..,M$ and $j=1,2,...,k$. with $N_{M,p}^{j}=N_{j}(T_{M,p})$. For each subject we also observe a $d\times 1$ vector of covariates $Z$. Now  we observe $n$ i.i.d. (independent and identically distributed) copies of  $ \{M,T_{M,p},N_{M,p}^{1},...,N_{M,p}^{k}, Z \}$, $p=1,2,...,M$. Accordingly,  observed data will be of the form $\{m_{i},t_{m_{i},p},n_{m_{i},p}^{1},...,n_{m_{i},p}^{k},z_i\}$, $p=1,2,...,m_{i}$ and $i=1,2,...,n$.

The regression analysis of panel count data based on maximum likelihood methods with a single failure mode is explored by Zhang (2002) and Wellner and Zhang (2007). Sreedevi and Sankaran (2020) studied panel count data with multiple modes of recurrence and developed a pseudo likelihood function for the observed data and derived an isotonic regression estimator (IRE) for cause specific mean functions.  They constructed a  pseudo likelihood function for the observed data (which does not involve covariates) by assuming that the successive counts of the recurrent event process $N_j(t)$ are independent random variables. We extend their derivation of pseudo likelihood function  into a scenario with covariates. We estimate both $\Lambda_{0j}(.)$ and $\beta_j$ simultaneously as the values that maximize the pseudo likelihood. Under the assumption that the underlying counting process $N_j(t)$ is a non-homogeneous Poisson process with conditional mean function given in Eq. (\ref{pmm}), for $j=1,2,..,k$, we obtain
\begin{equation}
P (N_j(t)=m \vert Z) =\frac{(\Lambda_{0j}\text{exp}(\beta_j^{'}Z))^{m}\text{exp}(\Lambda_{0j}\text{exp}(\beta_j^{'}Z))    }  {m!}~~~\text{for}~~~ m=0,1,2,....
\end{equation}

When the $k>1$ modes of recurrence are independent, the pseudo log likelihood function of the observed data can be written as
	\begin{equation}
	l_{n}(\beta_j,\underline \Lambda_{0j},\underline{X})= \sum\limits_{j=1}^{k}l_{nj}(\beta_j,\Lambda_{0j},\underline{X}),
	\end{equation}
	where $\underline{X}$ is the observed data given by $\underline{X}=\{M,T_{M,p},N_{M,p}^{1},...N_{M,p}^{k}, Z \}$ and $l_{nj}(\Lambda_{0j},\underline{X})$ is the log likelihood corresponding to $j$ th cause. By extending the results in Zhang (2002) for panel count data with single mode of failure,  after ignoring  the insignificant parts in the estimation of $\beta_j$'s and  $\Lambda_{0j}$'s, $\l_{nj}(\beta_j,\Lambda_{0j},\underline{X})$ is given by
	\begin{equation} \label{lnj}
	l_{nj}(\beta_j,\Lambda_{0j},\underline{X})=\sum\limits_{i=1}^{n}\sum\limits_{p=1}^{M_{i}}[ N^{j}_{M_{i},p} \text{log} \Lambda_{0j}(T_{M_{i},p}) + N^{j}_{M_{i},p}(\beta_j^{'}Z_i) -\Lambda_{0j}(T_{M_{i},p})\text{exp}(\beta_j^{'}Z_i)]~~~ \text{for}~~~ j=1,2,...,k
\end{equation}
where $M_{i}$ is the number of observation times, $T_{M_{i},p},~p=1,2,.., M_{i}$, the different observation times and $N^{j}_{M_{i},p},p=1,2,...,M_{i}, j=1,2,..,k$ the number of recurrences of the event due to cause $j$ for $i$ th individual. We assume that given the covariate vector $Z$, the the distributions of $T$ and $M$ are independent of $\beta_j$ and $\Lambda_{0j}$. We maximize the log pseudo likelihood given in Eq. (\ref{lnj}) to obtain the estimators of $\beta_j$ and $\Lambda_{0j}$.\\
Now we can discuss the computational procedures. Based on the observed data $\underline{X}$ discussed above, we define the following terms.  Let $I(A)$ be the indicator function of the set $A$ and $s_{1}<s_{2}<...<s_{r}$ be the distinct ordered observation time points in the set $\{T_{M_{i},p}, p=1,2,...,M_{i}, i =1,2,...,n$\}. For $ q \in \{1,2,...,r\}$  and for any particular cause of recurrence $J$, define
\begin{equation} \label{bqj}
b_{qj}=\sum\limits_{i=1}^{n}\sum\limits_{p=1}^{M_{i}} I[{T_{M_{i},p}}= s_{q};J=j ]
\end{equation}
the number of observations made at $s_{q}$ due to cause $j$ and
\begin{equation}\label{nbarqj}
\Bar{n}_{qj}= \frac{1}{b_{qj}}\ \sum\limits_{i=1}^{n}\sum\limits_{p=1}^{M_{i}}N^{j}_{M_{i},p}I[{T_{M_{i},p}}= s_{q};J=j ]
\end{equation}
as the mean value of the recurrences made at $s_{q}$ due to cause $j$ for $j=1,2,...,k$. Also define
\begin{equation}\label{vz}
V_{qj}(\beta_j,Z)= \frac{1}{b_{qj}}\sum\limits_{i=1}^{n}\sum\limits_{p=1}^{M_{i}}\text{exp}(\beta_j^{'}Z_i)I[{T_{M_{i},p}}= s_{q};J=j ]
\end{equation}
and
\begin{equation}\label{wz}
W_{qj}(\beta_j,Z, N)=\frac{1}{b_{qj}}\sum\limits_{i=1}^{n}\sum\limits_{p=1}^{M_{i}} {T_{M_{i},p}} (\beta_j^{'}Z_i)I[{T_{M_{i},p}}= s_{q};J=j ].
\end{equation}
Now  we can rewrite the log pseudo likelihood for $j$ th mode of recurrence given in Eq.(\ref{lnj}) as
\begin{equation}\label{newln}
l_{nj}(\beta_j,\Lambda_{0j} \vert \underline{X})= \sum\limits_{q=1}^{r} b_{qj}[\Bar{n}_{qj} \text{log}\Lambda_{0j}(s_{q})-V_{qj}(\beta_j,Z)\Lambda_{0j}(s_{q})+W_{qj}(\beta_j, Z,N_{j})] .
\end{equation}
We maximize Eq.(\ref{newln}) to obtain the estimates of $\beta_j$ and $\Lambda_j(.)$ for $j=1,2,...,k$. The obtained semiparametric maximum pseudo likelihood estimators will be the values of parameters that maximize
(\ref{newln}) over the set $R^{d} \times \Omega^{+}$ where $R$ is the set of real numbers and $\Omega^{+}=\{ (y_1,y_2,...,y_r) \in R^{d}:y_1 \le y_2 \le...\le y_r\}$. The estimators can be obtained as
\begin{equation}
(\widehat{\beta_j},\widehat{\Lambda_{0j}})= \underset{(\beta_j,\Lambda_{0j}) \in R^{d} \times \Omega^{+}}{\mathrm{arg max}}{l_{nj}(\beta_j,\Lambda_{0j} \vert \underline{X})}.
\end{equation}
To solve the optimisation problem numerically, we first choose an initial value of $\beta_j$, say $\beta_j^{0}$. Now for a fixed $\beta_j$, the estimator of $\Lambda_{0j}$ can be obtained as $\hat \Lambda_{0j}(\beta_j^{0})=\underset{\Lambda_{0j} \in \Omega^{+} }{\mathrm{arg max}}(l_{nj}^{*}(\Lambda_0j \vert \beta^{0}_{j}, \underline{X}))$  where
\begin{equation}\label{l*}
l_{nj}^{*}(\Lambda_0j \vert \beta_j, \underline{X})=\sum\limits_{q=1}^{r}b_{qj}[\Bar{n}_{qj}\text{log}\Lambda_{0j}(s_{q})-V_{qj}(\beta_j,Z)\Lambda_{0j}(s_{q})].
\end{equation}
Let $\Lambda_{0j}^{0}$ be the solution of Eq. (\ref{l*}). Now using the estimated value of $\Lambda_{0j}^{0}$, we can find the updated estimate of $\beta_j$ as $\hat \beta_j(\Lambda_{0j}^{0})=\underset{\beta_j \in R^{d} }{\mathrm{arg max}}(l_{nj}^{**}(\beta_j \vert \Lambda_{0j}^{0},\underline{X})$ where
\begin{equation}\label{l**}
l_{nj}^{**}(\beta_j\vert \Lambda_0j, \underline{X})=\sum\limits_{q=1}^{r}b_{qj}[W_{qj}(\beta_j, Z,N_{j})-V_{qj}(\beta_j,Z)\Lambda_{0j}(s_{q})].
\end{equation}
The process is continued until the estimators converge. The convergence criteria can be chosen as
\begin{equation}\label {cc}
\displaystyle\left\lvert \frac{l_{nj}^{(h+1)}-l_{nj}^{(h)}}{l_{nj}^{(h)}} \right \rvert \le\epsilon,
\end{equation}
where $l_{nj}^{(h)}=l_{n}(\beta_j^{(h)},\Lambda_{0j}^{(h)})$ for $h=0,1,2,....$.\\
To estimate $(\beta_j, \Lambda_{0j})$ for $j=1,2,...,k$, the computational algorithm can be summarised as follows
\vspace{-0.1cm}
\begin{enumerate}[Step 1:]
	\item Choose an initial value $\beta_j$ say $\beta_j^{0}$.
	\item For the given  $\beta_j^{h}$, compute $\Lambda_{0j}^{h}$ as the maximum argument of Eq. (\ref{l*}), given by
	\begin{equation*}
	    \hat \Lambda_{0j}^{h}(\beta_j^{h})=\underset{\Lambda_{0j} \in \Omega^{+} }{\mathrm{arg max}}(l_{nj}^{*}(\Lambda_0j \vert \beta^{h}_{j}, \underline{X}).
	\end{equation*}
	\item  Update the estimate of $\beta_j^{h}$, using the estimate of $\Lambda_{0j}^{h}$ obtained in Step 2, as the maximum argument of Eq. (\ref{l**}), given by
	\begin{equation*}
	   \hat \beta_j(\Lambda_{0j}^{h})=\underset{\beta_j \in R^{d} }{\mathrm{arg max}}(l_{nj}^{**}(\beta_j \vert \Lambda_{0j}^{h},\underline{X})
	\end{equation*}
	 and obtain the value of $\beta_{j}^{(h+1)}$.
	\item Repeat the steps 2 and 3 for $h=1,2,...$, until converge criteria in Eq. (\ref{cc}) obtained.
\end{enumerate}
\section{Asymptotic Results}
The asymptotic properties of the proposed estimators can be derived using results from empirical process theory. Zhang (2002) proved some results of about the asymptotic behaviour of the semiparametric pseudo maximum  likelihood estimators when only a single mode of recurrence is observed and later Wellner and Zhang (2007) modified the results. When recurrence due to multiple modes are observed, Sreedevi and Sankaran (2020) studied about the asymptotic properties of cause specific mean functions. We extend the results from Wellner and Zhang (2000) into a multiple cause scenario and generalize the results discussed in Sreedevi and Sankaran (2020) to incorporate covariates. We establish the asymptotic normality and strong consistency of the proposed estimators.

As we discuss, we estimate $\beta_j$ and $\Lambda_{0j}$ for $j=1,2,...,k$ as the maximum points of the pseudo likelihood function given in Eq.(\ref{lnj}). We assume that the estimators as well as the true value of the parameters include in the parameter domain $\mathcal{R} \times \mathcal{F}_j$ where $\mathcal{R} \in \mathcal{R}^d$ is a bounded convex set and  $\mathcal{F}_j$ be the class of functions defined as
\begin{equation*}
    \mathcal{F}_j \equiv \{\Lambda_j(.):[0,\infty) \to [0,\infty)|~~\Lambda_j(.)~~\text{is monotone non-decreasing with}~~ \Lambda_j(0)=0 \}~~~~~j=1,2,...,k.
\end{equation*} To prove the asymptotic properties of the estimators, we define the following. Let $\mathcal{B}_d$ and $\mathcal{B}$ denote the collection of Borel sets in $\mathcal{R}_d$ and $\mathcal{R}$ respectively. Let $H(.)$ be the distribution of the covariate vector $Z$ and $\tau$=max$(t)$ and define $\mathcal{B}_1[0,\tau]=\{B \cap [0,\tau]:B \in \mathcal{B}\}$ 
Now we define the measures $\psi_{j}$, $\eta_{j}$, and $\theta_j$ as follows. For $B, B_1 \in \mathcal{B}_1[0,\tau]$ and $C\in \mathcal{B}_d$ define
\begin{equation*}
    \eta_{j}(B\times C)=\int_C \sum\limits_{m=1}^{\infty}P(M=m;J=j \vert Z=z) \times\sum\limits_{p=1}^{m} P(T_{m,p} \in B \vert M=m, Z=z) dH(z).
\end{equation*}
A similar measure is defined by Shick and Yu (2000) to study the consistency of the likelihood estimators for mixed case interval censored data.
Define the $L_2$ metric $d_1(.)$ in parameter space $\mathcal{R} \times \mathcal{F}_j$ as
\begin{equation*}
    d_1((\beta_{j1},\Lambda_{0j1}),(\beta_{j2},\Lambda_{0j2}))=\{|\beta_{j1}-\beta_{j2}|^2+\| \Lambda_{0j1}-\Lambda_{0j2}\|^2_{L_2(\psi_{1j})}\} ^{\frac{1}{2}},
\end{equation*}
where $(\beta_{j1},\Lambda_{0j1})$ and  $(\beta_{j2},\Lambda_{0j2})$ are elements of the parameter space $R^{d} \times \Omega^{+}$
and
$\psi_{j}(B)=\eta_{j}(B \times \mathcal{R}^d)$.
 Now to establish the strong consistency of the estimators, we state the following regularity conditions.\\
\textbf{C1:} The true parameter values of $\beta_j$ and $\Lambda_{0j}$ include in $\mathcal{R}^{0} \times \mathcal{F}$, where $\mathcal{R}^{0}$ is the interior of $\mathcal{R}$.\\
\textbf{C2:} The observation times $T_{M,p}$ are the random variables included in the bounded interval $[0,\tau]$ for some $\tau \in (0,\infty)$ for all $p=1,2,...,M$, $M=1,2,...$. Also the measure $\psi_{j} \times H$ on $([0,\tau] \times\mathcal{R}^{d}, \mathcal{B}_{1}[0,\tau] \times \mathcal{B}_{d})$ is absolutely continuous with respect to $\eta_{j}$ for  $j=1,2,...,k$ and $E(M) < \infty $.\\
\textbf{C3:} For each the true baseline cumulative mean function $\Lambda_{0j}$,$j=1,2,...,k$ , there exist and $I_j \in (0,\infty) $ such that $\Lambda_{0j}(\tau) \le I_j$.\\
\textbf{C4:} The function $I_{0j}$ defined as $I_{0j}(X) \equiv \sum\limits_{p=1}^{M}N_{M,p}\text{log}(N_{M,p})$, satisfies $P(I_{0j}(X)) <\infty$.\\
\textbf{C5:} The support of $H$, the distribution of covariate vector $Z$ is a bounded set in $\mathcal{R}^{d}$.\\
\textbf{C6:} For all $a \in \mathcal{R}^{d}, a \neq 0$ and $c \in \mathcal{R}$, $P(a^{'}Z \neq c) >0$.\\
\textbf{ Theorem 1}\\
Under the above regularity conditions C1-C6, and the proposed model specified by Eq. (\ref{pmm}), for every $b<\tau$ such that $\psi_{1j}([b,\tau])>0$,
\begin{equation*}
    d_{1}((\hat\beta_j,\hat \Lambda_{0j}I_{[0,b]}),(\beta_j, \Lambda_{0j}I_{[0,b]})) \to 0~~~~~~~\text{a.s}~~~~~n \to \infty.
\end{equation*}
Specifically , when $\psi_{j}({\tau}) >0$, we have
\begin{equation*}
    d_{1}((\hat\beta_j,\hat \Lambda_{0j}),(\beta_j, \Lambda_{0j})) \to 0~~~~~~~\text{a.s}~~~~~n \to \infty.
\end{equation*}
The proof of the result can be derived by extending the results in Wellner and Zhang (2007).\\
To derive the rate of convergence, apart from the above stated regularity conditions, we also suppose that \\
\textbf{C7:} For some interval $O_{j}[T]=[\sigma_j,\tau]$ with some $\sigma_j >0$ with $\Lambda_{0j}(\sigma_j) >0 $ and $P({\bigcap}_{p=1}^M){T_{M,p} \in [\sigma_j,\tau]})=1$.\\
\textbf{C8:} The number of observations are bounded ie. $P(M \le m_0)=1$ for some $m_0 \le \infty$.\\
\textbf{C9:} For some $z_{0j} \in (0,\infty)$ the function $Z \to E (e^{z_{0j}N_j(\tau)})$ is uniformly bounded for any $Z$ and for all $j=1,2,..,k$.\\
\textbf{C10:} There exists a constant $s_0 >0$ such that $P(T_{M,p}-T_{M,p-1} \ge s_0\text{for all}~~ p=1,2,...,M$)=1. Also, $\psi_{j}(t)$ is absolutely continuous with respect to a Lebesgue measure with $0<c_{0}< \psi_{j}^{'}(t)$ where $c_{0j}$ is a positive constant and $\psi_{j}^{'}$ is the derivative of $\psi_{j}$.\\
\textbf{C11} : The true baseline cumulative  mean functions $\Lambda_{0j}$'s are differentiable and the derivatives has positive and finite lower and upper bounds in the observation interval for all $j=1,2,..,k$. ie for each $j$ there exists a constant $l_j$ such that $\frac{1}{l_j} \le \Lambda_{0j}^{'}(t) \le l_j \le \infty$ for $t \in O[T]$.\\
\textbf{Theorem 2:} \\Under the above stated regularity conditions C7-C11 and the conditions C1-C6 stated to prove consistency of the estimators, for the constant $z_{0j}$ defined in C9, satisfying $z_{0j} \ge 4m_0(1+g_{0j})^{2}$ where 
$g_{0j}= \sqrt{sign{c_{0j}\Lambda_{0j}^{3}(\sigma_j)/(24.8l_j)}}$ and $\psi_{j}({\tau})>0$,
\begin{equation*}
    n^{\frac{1}{3}}d_{1}(\hat\beta_j,\hat \Lambda_{0j}),(\beta_j, \Lambda_{0j})=O_p(1).
\end{equation*}
We can see that the rate of convergence of estimators is of order  $n^{-\frac{1}{3}}$ only. Even though the over all convergence rate is $n^{-\frac{1}{3}}$, we can establish the asymptotic normality of regression parameters, with the rate of convergence $n^{-\frac{1}{2}}$. Huang (1996) considered this similar situation for current status data and Sreedevi et al. (2017) proved similar results for current status data with competing risks. \\
\textbf{ Theorem 3:} \\Under the regularity conditions for Theorem 2, the estimator $\hat \beta_j$ is asymptotically normal  and
\begin{equation*}
    n^{\frac{1}{2}}(\hat \beta_j-\beta) \to_{d} \bar Z,
\end{equation*}
\vspace{-0.1cm}
where $\bar Z \sim N_{d}(0, \Sigma ^{-1}\Theta (\Sigma ^{-1})^{'})$ with
\begin{equation*}
  \Theta=E \Big (\sum\limits_{p,p^{'}=1}^{M} C^{j}_{p,p^{'}}(Z)[Z-R(M,T_{M,p})][Z-R(M,T_{M,p})]^{'}\Big ),
\end{equation*}
\begin{equation*}
  \Sigma=E \Big (\sum\limits_{p=1}^{M} \Lambda_{j}(T_{M,p})exp(\beta_j^{'}Z) [Z-R(M,T_{M,p})]^{\bigotimes2}\Big ),
\end{equation*}
in which, $R(M, T_{M,p}) \equiv E(Z\text{exp}(\beta_j^{'}Z) \vert M, T_{M,p})/E(\text{exp}(\beta_j^{'}Z) \vert M, T_{M,p})$ and \\ $C^{j}_{p,p^{'}}(Z)=\text{Cov}(N_{j}(T_{M,p}),N_{j}(T{M,p^{'}}))$. We can see that, in general $\hat\beta_j$ is not asymptotically efficient, but when the counts $\{N_j(T_{M,p}),p=1,2,...,M\}$ consist a cluster of Poisson count data where the counts within the cluster are independent, the estimator $\beta_j$ become asymptotically efficient for $j=,1,...,k$. Proof of Theorem 3 can be obtained as a generalisation of the results in Wellner and Zhang (2007).

\section{Simulation Study}
We carry out a Monte Carlo simulation study to assess the performance of the proposed estimation procedure in finite samples. We consider the situation with two competing risks.
The real life situations in reliability and survival studies are taken as a model to generate panel count data of the form
$ \{M_{i},T_{M_{i},p},N_{M_{i},p}^{1},N_{M_{i},p}^{2},Z_i\}$ for $p=1,2,...,M_{i}$, $i=1,2,...,n$. We consider $Z_i=\{Z_{i1},Z_{i2}\}^{'}$, as the covariate vector with two mutually independent components. For each subject,  $Z_{i1}$ is generated from a Bernoulli distribution with probability of success $0.5$ and $Z_{i2}$ is generated from a Normal distribution with mean $0$ and standard deviation $0.5$ .
The number of observation times $M_{i}$ for each individual is generated from a discrete uniform distribution $U(1,5)$ for $i=1,2,...,n$. Thus the maximum number of observations for each individual is restricted upto 5. Then we generated gap times between each observation from uniform distribution $U(1,5)$. The discrete observation time points $T_{M_{i},p}$ for $p=1,2,...,M_i$ and $i=1,2,...,n$ are generated using the above mentioned time gaps. Once the observation times are generated, number of recurrences $\{N_{M_{i},p}^{1},N_{M_{i},p}^{2}\}$ are generated from a bivariate Poisson process given by
\begin{equation}
(\Delta N_{M_i,p}^{1},\Delta N_{M_i,p}^{2}) \sim \text{BivPo}(\Lambda_{01}(\Delta T_{M_i,p})\text{exp}(\beta_{1}^{'}Z_i),\Lambda_{02}(\Delta T_{M_i,p})\text{exp}(\beta_{2}^{'}Z_i),\rho),
\end{equation}
where $ \Delta N_{M_i,p}^{j}= N_{M_i,p}^{j}- N_{M_i,p-1}^{j}$ for $j=1,2$;
$\Delta T_{M_i,p}=T_{M_i,p}-T_{M_i,p-1}$, $\Lambda_{01}(t)$ and $\Lambda_{02}(t)$ are the true baseline functions due to mode l and mode 2, $\beta_{1}$ and $\beta_{2}$ are the values of regression parameters due to mode l and mode 2, and $\rho$ is the covariance between the number of recurrences due to mode 1 and mode 2.

We consider two different forms of $\Lambda_{01}(t)$ and $\Lambda_{02}(t)$ , $t$ and  $2t$ to generate panel count data.
The sample size $n$ takes three different values $n=50,100,200$.  The process is repeated 10000 times to estimate the efficiency of the estimators. The absolute bias and mean square error (MSE) of the estimates of $\beta_{1}=\{\beta_{11},\beta_{12}\}$ and $\beta_{2}=\{\beta_{21},\beta_{22}\}$  are obtained.

Various parameter values of $\beta_{1}$ and $\beta_{2}$ are considered. Since the results are similar we present the same only for three different combinations of $\beta_{1}$ for $\beta_{2}$ in Tables 1-3. To obtain the convergence we choose $\epsilon=10^{-5}$. The covariance $\rho$ is set to be $0.5$ in our studies. The simulations are carried out using R programming language.
\begin{table}[ht]
	\caption{ Absolute bias and MSE of the estimators of regression coefficients}
	\begin{small}
		\centering
		\begin{tabular}{|c|r|rr rr|rr rr|}
			\hline
			\cline{3-10}&&\multicolumn{4}{c|}{$(\beta_{11},\beta_{12})=(0.5,1)$}&
			\multicolumn{4}{c|}{$(\beta_{21},\beta_{22})=(-1,0.5)$}
		\\
			\cline{3-10} True Baseline Function  & $ n$
			&Bias11&Bias12&MSE11&MSE12&Bias21&Bias22&MSE21&MSE22\\
			\hline
			\multirow{3}{*} {$\Lambda_{01}(t)=t,\Lambda_{02}(t)=2t$}& 50& 0.0248&0.0382&0.0218&0.0108&0.1210&0.0216&0.1080& 0.0343\\
			& 100 & 0.0127&0.0137&0.0099&0.0093&0.0928&0.0199 &0.0856&0.0243
			\\
			& 200 & 0.0098&0.0117&0.0012&0.0065&0.0720 &0.0076 &0.0098&0.0105
			\\
			\hline
			\multirow{3}{*}{$\Lambda_{01}(t)=2t,\Lambda_{02}(t)=2t$} & 50 & 0.0198& 0.0454& 0.0194&0.0121&0.0211&0.0278& 0.1097&0.0218\\
			& 100 & 0.0114&0.0218  &0.0089&0.0074&0.0141& 0.0124&0.0954&0.0122
			\\
			& 200 & 0.0073&0.0098&0.0065&0.0059&0.0069&0.0072&0.0088&0.0076
			\\
			\hline
			\multirow{3}{*}{$\Lambda_{01}(t)=t,\Lambda_{02}(t)=t$} & 50& 0.0132& 0.0245& 0.0110& 0.0279& 0.0510&0.0199& 0.0350&0.0214 \\
			& 100 & 0.0093&0.0135&0.0065& 0.0138& 0.0061& 0.0131& 0.0102& 0.0166
			\\
			& 200 &0.0043 &0.0089 &0.0027&0.0020& 0.0045& 0.0091&0.0071&0.0065
			\\
			\hline
		\end{tabular}
	\end{small}
	\label{31}%
\end{table}%
\begin{table}[h]
	\caption{ Absolute bias and MSE of the estimators of regression coefficients}
	\begin{small}
		\centering
		\begin{tabular}{|c|r|rr rr|rr rr|}
			\hline
			\cline{3-10}&&\multicolumn{4}{c|}{$(\beta_{11},\beta_{12})=(1,0.5)$}&
			\multicolumn{4}{c|}{$(\beta_{21},\beta_{22})=(0.5,1)$}
		\\
			\cline{3-10} True Baseline Function  & $ n$
			&Bias11&Bias12&MSE11&MSE12&Bias21&Bias22&MSE21&MSE22\\
			\hline
			\multirow{3}{*} {$\Lambda_{01}(t)=t,\Lambda_{02}(t)=2t$}& 50& 0.1311&0.0393&0.0198&0.0187&0.0312&0.0247&0.1201& 0.0298\\
			& 100 &0.1008&0.0137&0.0108&0.0082&0.0219&0.0187 &0.0916&0.0117
			\\
			& 200 & 0.0832&0.0121&0.0084&0.0051&0.0119 &0.0931 &0.0411&0.0095
			\\
			\hline
			\multirow{3}{*}{$\Lambda_{01}(t)=2t,\Lambda_{02}(t)=2t$} & 50 & 0.0278& 0.0298& 0.0171&0.0186&0.0217&0.0354& 0.0521&0.0221\\
			& 100 & 0.0211&0.0218  &0.0092&0.0110&0.0156& 0.0219&0.0954&0.0131
			\\
			& 200 & 0.0102&0.0102&0.0059&0.0072&0.0091&0.0141&0.0127&0.0091
			\\
			\hline
			\multirow{3}{*}{$\Lambda_{01}(t)=t,\Lambda_{02}(t)=t$} & 50& 0.0538& 0.0213& 0.0194& 0.0232& 0.0144&0.0213& 0.0212&0.0273 \\
			& 100 & 0.0391&0.0104&0.0102& 0.0117& 0.0081& 0.0121& 0.0126& 0.0139
			\\
			& 200 &0.0708 &0.0091 &0.0083&0.0054& 0.0042& 0.0072&0.0087&0.0076
			\\
			\hline
		\end{tabular}
	\end{small}
	\label{31}%
\end{table}%
\begin{table}[ht]
	\caption{ Absolute bias and MSE of the estimators of regression coefficients}
	\begin{small}
		\centering
		\begin{tabular}{|c|r|rr rr|rr rr|}
			\hline
			\cline{3-10}&&\multicolumn{4}{c|}{$(\beta_{11},\beta_{12})=(1,-2)$}&
			\multicolumn{4}{c|}{$(\beta_{21},\beta_{22})=(-1,2)$}
		\\
			\cline{3-10} True Baseline Function  & $ n$
			&Bias11&Bias12&MSE11&MSE12&Bias21&Bias22&MSE21&MSE22\\
			\hline
			\multirow{3}{*} {$\Lambda_{01}(t)=t,\Lambda_{02}(t)=2t$}& 50& 0.0384&0.0421&0.0212&0.0186&0.0492&0.0291&0.0293& 0.0418\\
			& 100 & 0.0276&0.0187&0.0109&0.0115&0.0321&0.0182&0.0172&0.0329
			\\
			& 200 & 0.0119&0.0113&0.0083&0.0085&0.0238 &0.0732 &0.0093&0.0228
			\\
			\hline
			\multirow{3}{*}{$\Lambda_{01}(t)=2t,\Lambda_{02}(t)=2t$} & 50 & 0.0275& 0.0471& 0.0256&0.0219&0.0221&0.0269& 0.0421&0.0253\\
			& 100 & 0.0262&0.0218  &0.0192&0.0143&0.0162& 0.0182&0.0321&0.0187
			\\
			& 200 & 0.0133&0.0128&0.0102&0.0092&0.0101&0.0116&0.0192&0.0092
			\\
			\hline
			\multirow{3}{*}{$\Lambda_{01}(t)=t,\Lambda_{02}(t)=t$} & 50& 0.0218& 0.0291& 0.0172& 0.0267& 0.0279&0.0199& 0.0401&0.0271 \\
			& 100 & 0.0113&0.0173&0.0108& 0.0129& 0.0162& 0.0131& 0.0284& 0.0192
			\\
			& 200 &0.0095 &0.0121 &0.0072&0.0081& 0.0119& 0.0091&0.0172&0.0125
			\\
			\hline
		\end{tabular}
	\end{small}
	\label{31}%
\end{table}%
From simulation studies, we observe that the absolute bias and MSE of the estimators of regression coefficients  approaches  zero as sample size increases. This ensure that the proposed estimators are unbiased with nominal variance. The minimum value of bias and MSE are obtained  when the true base line function takes the form $\Lambda_{0j}(t)=t$ for both $j=1,2$.
\section{Data Analysis}
The proposed estimation procedure is applied to a real data on  skin cancer chemoprevention trial given in Sun and Zhao (2013) for illustration. The primary objective of this study was to evaluate the effectiveness of the drug DFMO (difluoromethylornithine) in reducing new skin cancers in a population with a history of non-melanoma skin cancers, basal cell carcinoma and squamous cell carcinoma. The patients were randomly assigned into two groups a treatment group with oral DFMO at a daily dose 0.5 gm and a palcebo group with a matching dosage. The data consist of the details of 290 patients with history of non-melanoma skin cancers who were supposed to be assessed or observed every 6 months. However, the real observation and follow up times differ from patient to patient. The data include the number of recurrences of two types of recurrent events, basal cell carcinoma (BC) and squamous cell carcinoma (SC). We treat these two types of cancers as two modes of recurrence following Sreedevi and Sankaran (2020).

In the data set, the number of observations on an individual varies from 1 to 17 and time of observation varies from 12 to 1766 days. For each individual, the information on age, gender, DFMO status, number of prior tumours are observed. We consider all 290  patients in our analysis which include  174 are males and 116 female.  To obtain more explicit conclusions, we analyse the data on males and females separately by taking the covariate information on DFMO status and number of prior tumours. Out of 290 patients ,147 were assigned to the placebo group and reaming 147 were treated with oral DFMO. The number of prior tumours varies from 1 to 35. The estimates of regression parameters with corresponding  standard errors for males are given in Table 4.
\begin{table}[h!]
		\centering
	\caption{ Estimates of the regression parameters with corresponding standard error for males}
	\begin{small}
				\begin{center}
			\begin{tabular}{ |c|c|c|c|c| }
				\hline
				Cause & Covariate & Coefficient & SE & P-value\\
				
				\hline
				\hline
				BC&  DFMO  & -0.3715 & 0.2331& 0.0111\\
				& Prior tumours &0.0685&0.0103& 0.0005\\
				\hline
				SC & DFMO  & -0.2408 & 0.0460&0.0600\\
				& Prior tumours&0.1013&0.0292&0.0005\\
				\hline
			\end{tabular}
		\end{center}
	\end{small}
\end{table}
The baseline cause specific cumulative mean functions for males are plotted in Figure 1. The solid line represents the baseline cumulative mean function for patients with BC and dotted line represents the baseline cumulative mean function for patients with SC in Figure 1 and 2.
\begin{figure}[hbt!]
\begin{center}
    \includegraphics[height=10cm,width=120mm]{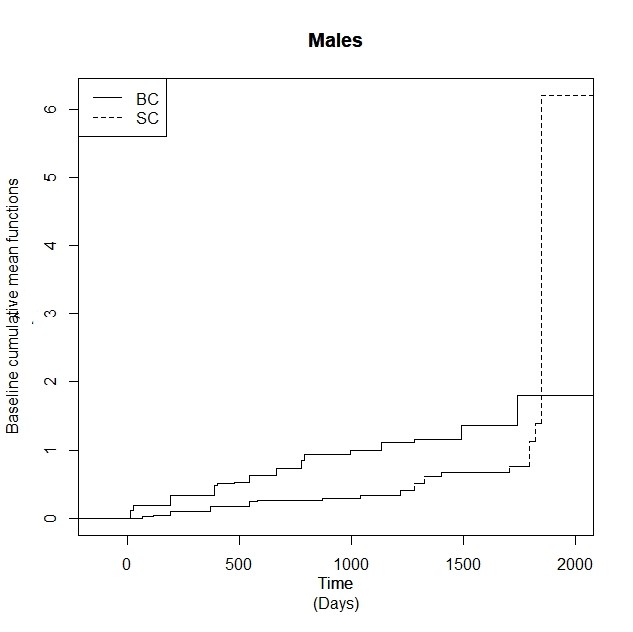}
	\caption{ Baseline cause specific cumulative mean functions for males }
\end{center}
\end{figure}
The estimates of regression parameters for females are given in Table 5 and the baseline cause specific cumulative mean functions are plotted in Figure 2.
\begin{table}[h!]
		\centering
	\caption{ Estimates of the regression parameters with corresponding standard error for females}
	\begin{small}
				\begin{center}
			\begin{tabular}{ |c|c|c|c|c| }
				\hline
				Cause & Covariate & Coefficient & SE & P-value\\
				
				\hline
				\hline
				BC&  DFMO  & -0.1671 & 0.0347& 0.0632\\
				& Prior tumours &0.0666&0.0456& 0.0033\\
				\hline
				SC & DFMO  & 0.9557 & 0.0964&0.0320\\
				& Prior tumours&0.1053&0.0458&0.0023\\
				\hline
			\end{tabular}
		\end{center}
	\end{small}
\end{table}

\begin{figure}[hbt!]
\begin{center}
    \includegraphics[height=10cm,width=120mm]{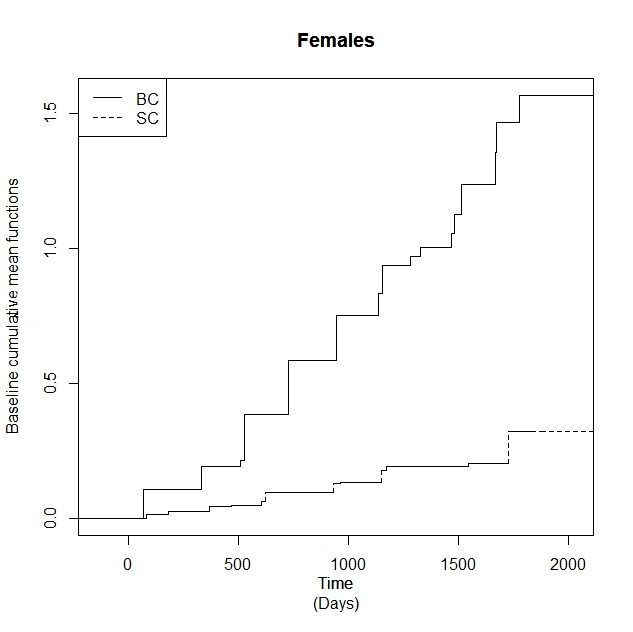}
	\caption{ Baseline cause specific cumulative mean functions for females }
\end{center}
\end{figure}
From Tables 4 and 5  we can see that modes of cancer recurrences basal cell carcinoma (BC) and squamous cell carcinoma (SC) affect males and females in different ways.  The regression estimators for number of prior tumours is grater than zero for both males and females and for both modes BC and SC. This implies that as the number of prior tumours increases hazard rate increases always. Since the hazard ratio is less than unity, we can say that the drug DFMO decreases the hazard rate for males with both BC and SC and for females with BC.

From the plots of baseline cumulative mean functions, we can see that recurrence rate of BC is higher in males than the recurrence rate of SC upto 1800 days (approximately) and from that point recurrence rate of SC crosses that of BC, while for females recurrence rate of SC is always lower than that of BC. The plots also show the difference in recurrence patterns of the events due BC and SC for males and females.
\section{Conclusion}
Panel count data with multiple modes of failure often arise in periodic follow up studies that consider recurrent events exposed to multiple modes. In this article, we proposed a new proportional mean model for the analysis of panel count data with multiple modes of recurrence. Estimators for regression parameters and baseline cumulative mean functions due to each recurrence mode are derived. A simple iterative procedure is developed for the estimation of parameters. The finite sample performance of the estimators in terms of bias and MSE is assessed through a Monte Carlo simulation study. A real data set on skin cancer chemo prevention trial is analysed using the proposed procedures.

The estimation procedure we developed in this article considered the pseudo likelihood function of panel count data. Maximum likelihood estimators in this situation can be developed by extending the results in Wellner and Zhang (2007), which involves a more complex iterative procedure. An approach based on estimating equations can also be examined for the regression analysis of panel count data with multiple recurrence modes. In many situations, rate functions of the underlying recurrent event process are of importance than mean functions. Cause specific rate functions developed by Sankaran et al. (2021) can be used to study panel count data, when subjects are exposed to multiple recurrence modes.

\end{document}